\documentstyle {article}

\begin{document}

{\bf Phase coherence appearance in thin superconducting film with strong
disorder. The return to the Mendelssohn model.}

\

A.V.Nikulov, Institute of Microelectronics Technology and High Purity
Materials, Russian Academy of Sciences, 142432 Chernogolovka, Moscow
District, Russia.

\

\begin{abstract} It is shown that transition from the mixed state without
the phase coherence into the mixed state with the long-range phase
coherence (i.e. into the Abrikosov state) of superconductors without
disorder must be first order phase transition. The phase coherence
appearance in thin superconducting film with strong disorder is considered.
The observed smooth transition is explained by increasing of the effective
fluctuation dimensionality (from zero to one) in superconductors with
strong disorder. Mendelssohn model is used for the explanation of the
resistive properties of films with strong disorder.  \end{abstract}


\section{Introduction}

	Investigation of the fluctuations have change the habitual notion about
nature of the mixed state of type II superconductors. The investigation of
conventional superconductors has shown already that the habitual notion
based on the results of the mean field approximation is no quite right. But
results of this investigation remain no enough widely known for the
present. Therefore the most scientists connect the change of the habitual
notion with the fluctuation investigation in high-Tc superconductors (HTSC)
\cite{blatter}.

	Now there is no reason to think that the fluctuation effects in HTSC
differ qualitatively from the one in conventional superconductors. Moreover
the comparison \cite{nik96} shows that many results obtained at the HTSC
investigation repeat the one obtained at the investigation of conventional
superconductors. But these results were interpreted by different way at
HTSC and conventional superconductors investigation.

	A sharp change of the resistive properties is observed below the second
critical field $H_{c2}$ both in bulk conventional superconductors
\cite{nik81l} and in $YBa_{2}Cu_{3}O_{7-x}$ single crystals \cite{welp}
with weak disorder. This sharp change was interpreted in \cite{nik81l} as a
transition from the mixed state without phase coherence (called as "one-
dimensional" state in this paper) into the Abrikosov state. According to
this interpretation the Abrikosov state is the mixed state with long-rang
phase coherence. The Abrikosov vortexes are singularity in the mixed state
with long-rang phase coherence. The Abrikosov vortexes appear in type II
superconductor because a magnetic flux can not be inside a superconducting
region with phase coherence and without singularity \cite{nik97} (the
Meissner effect). Consequently, the existence of the Abrikosov vortexes is
evidence of the long-rang phase coherence.

	According to the mean field approximation \cite{abrikos} the long-rang
phase coherence appears at $H_{c2}$ simultaneously with non-zero
superconducting electron density $n_{s} = |\Psi|^{2}$. And because the
magnetic field can not be inside a superconducting region with long-rang
phase coherence and without singularity \cite{nik97} the Abrikosov vortexes
appear at $H_{c2}$ also. This transition into the Abrikosov state was
considered as second order phase transition \cite{degennes}. This opinion
can be right in the mean field approximation but it can not be right in the
fluctuation theory, because the effective dimensionality of the fluctuation
decreases on two near $H_{c2}$ (in the lowest Landau level (LLL)
approximation region) \cite{lee72}.

	According to the mean field approximation the mixed state with long-rang
phase coherence (i.e. the Abrikosov state) can exist only. But according to
the fluctuation theory the mixed state without long-rang phase coherence
can exist also. According to the fluctuation theory the density of the
superconducting electrons above $H_{c2}$ is not equal zero and the length
of the phase coherence is small. Across a magnetic field direction it is
equal approximately $(\Phi_{0}/H)^{0.5}$. Where $\Phi_{0}$ is the flux
quanta, H is the magnetic field value. This state I call as the mixed state
without the phase coherence. The length of the phase coherence in the mixed
state without the long-rang phase coherence can be defined from the
correlation function $g(R) = <\Psi^{*}(R')\Psi(R'+R)>$.

	The correlation function of three-dimensional superconductors in a high
magnetic field (in the LLL region) in the linear approximation is
\cite{tinkha75}

$$g(R) = g(r,z) = A_{H}\exp(-|z|/\xi_{z})\exp(-r^{2}/\xi_{r}^{2})
\eqno{(1)}$$

at $|z| \gg \xi_{z}$ and $r \gg \xi_{r}$. Where $A_{H}$ is a coefficient;
$r = (x^{2} + y^{2})^{0.5}$; $\xi_{z} = (\Phi_{0}/2\pi (H-H_{c2})^{0.5}$;
$\xi_{r} = (2\Phi_{0}/\pi H)^{0.5}$. A magnetic field is parallel to z.
$\xi_{z}$ and $\xi_{r}$ are the length on which the phase coherence is
exist (i.e. $\xi_{z}$ and $\xi_{r}$ are the coherence length). The
relation for the longitudinal coherence length $\xi_{z} = (\Phi_{0}/2\pi
(H-H_{c2})^{0.5}$ can be valid in the linear approximation only and must be
renormalized near $H_{c2}$. Whereas the transversal coherence length
$\xi_{r} = (2\Phi_{0}/\pi H)^{0.5}$ changes little near $H_{c2}$ and
$\xi_{r}$ value is close to $(\Phi_{0}/H)^{0.5}$ in the whole LLL region.
This means that the correlation function of bulk superconductor in the LLL
region is similar to the one of one-dimensional superconductor. The
correlation function at $H_{c2}$ differs qualitatively from the one in zero
magnetic field. All components of the coherence length, i.e. $\xi_{x}$,
$\xi_{y}$ and $\xi_{z}$, increase up to the infinity at the critical
temperature where the second order phase transition takes place.

	Thus, if we define the phase coherence by the correlation function we
can conclude that the long-rang phase coherence can not be in the mixed
state of type II superconductors. On other hand we know that the Abrikosov
state is the mixed state with long-rang phase coherence. Consequently, the
definition of the phase coherence by the correlation function is unsuited
for the mixed state. It must follow from a right definition that the
existence of the singularities (the Abrikosov vortexes) is evidence of the
phase coherence.

	The Meissner effect is the consequence of the relation \cite{huebener}

$$\frac{\Phi_{0}}{2\pi}\int_{l}dR \frac{d\phi}{dR} =
\int_{l}dR\lambda_{L}^{2}j_{s} + \Phi \eqno{(2)}$$

where l is a closed path of integration; $\lambda_{L} =
(mc/e^{2}n_{s})^{0.5}$ is the London penetration depth; $j_{s}$ is the
superconducting current density; $\Phi$ is the magnetic flux contained
within the closed path of integration l. If the singularity is absent
$\int_{l}dR d\phi /dR = 0$. In this case the relation (2) is the equation
postulated by F. and H.London \cite{london35} for the explanation of the
Meissner effect \cite{meissner} (see \cite{huebener}). In the Abrikosov
state $(\int_{l}dR d\phi/dR)/2\pi = n$ is a number of the Abrikosov vortex
contained within the closed path of integration. The relation (2) is
gauge-invariant. It is valid if the phase coherence exists along the closed
path of integration. Consequently, we can use the relation (2) for the
definition of the phase coherence: the phase coherence exists in some
region if the relation (2) is valid for any closed path in this region. It
is obvious that according to this definition the long-rang phase coherence
exists both in the Meissner and Abrikosov states. If the closed path is
great (i.e. if it's radius $\gg \lambda_{L}$) then
$\int_{l}dR\lambda_{L}^{2}j_{s}$ value is small in comparison with other
terms of (2). In this case the condition of the phase coherence is
$(\int_{l}dR d\phi/dR)/2\pi = \Phi/\Phi_{0} = n$.

	In the mixed state without the phase coherence the relation (2)
is valid for the closed path of integration l with a radius $r =
l/2\pi < (\Phi_{0}/H)^{0.5}$ only (across magnetic field). Therefore, in
this case the magnetic field can penetrate in a superconductor without
singularities (i.e. without Abrikosov vortexes) as well as in the normal
state. Because in the mixed state without the phase coherence the
transversal length of the phase coherence $\xi_{r} \simeq
(\Phi_{0}/H)^{0.5}$ changes little near $H_{c2}$ two characteristic lengths
only ($(\Phi_{0}/H)^{0.5}$ and sample size L) are across magnetic field
direction in a superconductor without disorder. Therefore the long-rang
phase coherence appearance must be a first order phase transition in ideal
superconductors. The length of the phase coherence must change by jump from
$(\Phi_{0}/H)^{0.5}$ to L at this transition.

	The resistivity has different nature in the mixed states with and
without the phase coherence. The cause of the resistivity in the mixed
state without the phase coherence does not differ qualitatively from the
one in the normal state. The resistivity value is decreased in the
consequence of the paraconductivity \cite{nik81l}. Whereas, the voltage can
be in the Abrikosov state only if the phase difference changes in time.
The change of the phase difference in time in the Abrikosov state means the
vortex flow. The resistivity caused by the vortex flow is called "flux flow
resistivity" \cite{huebener}. But it is no quite right denomination because
the magnetic flux does not flow. I will use more right denomination "vortex
flow resistivity" instead of "flux flow resistivity". I will considered the
resistivity in the mixed state without the phase coherence as the
paraconductivity regime and in the Abrikosov state as the vortex flow
regime.

	The vortex flow is impeded by disorder of the superconductor. In the
consequence of this the current-voltage characteristics of superconductor
in the Abrikosov state can be non-Ohmic. And in some cases the resistivity
can be equal zero below the critical current \cite{huebener}. This
phenomena is called as vortex pinning effect \cite{campbell}. It is obvious
that the vortex pinning can not be in the state without the phase coherence
because the vortex (as the singularity in the state with the phase
coherence) can not exist without the phase coherence. Therefore the vortex
pinning effect may be considered as the consequence of the phase coherence.

	Consequently, a change of the resistive properties must be observed
first of all at the long-rang phase coherence appearance. The vortex
pinning appears and a transition from the paraconductivity regime to the
vortex flow regime takes place there. This change must be sharp in the ideal
superconductor. Therefore it is obvious that the sharp change of the
resistive properties observed in bulk superconductors with weak disorder
\cite{nik81l} at $H_{c4} < H_{c2}$ is the transition into the Abrikosov
state, because the vortex pinning appears and the vortex flow resistivity
decreases sharply at $H_{c4}$. (The designation $H_{c4}$ for the position
of the transition into the Abrikosov state was proposed in my paper
\cite{nik90}). The observation of the sharp change in \cite{nik81l} means
that the transition into the Abrikosov state in bulk superconductors with
weak disorder can be close to the one in the ideal superconductor. The
length of the phase coherence changes abruptly from $(\Phi_{0}/H)^{0.5} =
10^{-5} - 10^{-6} cm$ to a sample size $L = 0.01 - 1 cm$ in real cases.

	The same transition is observed in bulk conventional superconductor
\cite{nik81l} and in $YBa_{2}Cu_{3}O_{7-x}$ \cite{welp} (it was shown in
\cite{nik96} for example). But a conception of the vortex lattice melting
has become very popular at the HTSC investigation \cite{bishop}. Therefore
the sharp change of the resistive properties (kink) in
$YBa_{2}Cu_{3}O_{7-x}$ was interpreted in \cite{welp} and other papers as
the vortex lattice melting. This interpretation is very popular, but it can
not be right.

	According to this interpretation the transition from the vortex lattice
state into the vortex liquid state is observed at $H_{c4}$.  (Because the
same transition is observed in \cite{nik81l} and in \cite{welp} I use the
designation $H_{c4}$ in both cases). But the vortex liquid is the mixed
state with long-rang phase coherence (i.e. the Abrikosov state), because
the existence of the vortexes is evidence of the long-rang phase coherence.
According to the right definition by the relation (2) the existence of the
long-rang phase coherence does not depend on the crystalline long-rang
order of the vortex. Consequently, the first order phase transition from
the vortex liquid into the mixed state without the phase coherence must be
observed above the vortex lattice melting in ideal superconductor. The
sharp change of the resistive properties must be observed at this
transition. But no sharp change is observed above $H_{c4}$ both in
conventional superconductors \cite{nik81l} and in $YBa_{2}Cu_{3}O_{7-x}$
\cite{welp}. Consequently, the transition from the mixed state without
the phase coherence into the Abrikosov state is observed both in
\cite{nik81l} and in \cite{welp}. Result of \cite{shilling} shows that this
transition in bulk superconductors with weak disorder can be first order
indeed.

	It is proposed in vortex lattice melting theories that the Abrikosov
state is the vortex lattice. This opinion is based on the Abrikosov
solution \cite{abrikos}. But it not proved that this solution is valid in
thermodynamic limit. It must be emphasized that the direct observations
(\cite{troible} and many others) do not prove the validity of the Abrikosov
solution \cite{abrikos}. The experimental results are obtained by the
investigation of real samples whereas the Abrikosov solution is obtained
for the ideal case. Larkin \cite{larkin70} has shown that the crystalline
long-rang order of the vortex lattice is unstable against the introduction
of random pinning. Consequently, the crystalline long-rang order does not
exist in real superconductors and can be in the ideal case only. But the
existence of the Abrikosov state in the ideal case has not been
proved for the present because the mean field approximation is not valid in
this case.

	Therefore, the vortex lattice melting theories are unsatisfactory
in principle, since they start from the state in which the translation
symmetry has been broken by hand \cite{tesan91}. Some theorists consider no
the vortex lattice melting but the solidification transition of vortexes
\cite{macikeda, moore}. They do not propose the Abrikosov state existence,
but try to find the transition to it. But the phase coherence is defined by
the correlation function in the solidification theory. According to this
definition the long-rang phase coherence can not exist without the
crystalline long-rang order of the vortexes. It is claimed in some works
\cite{ikeda} that the phase coherence can remain short-ranged even in the
vortex solid phase. Therefore the solidification theories \cite{macikeda,
moore} find the transition into the Abrikosov state as the vortex lattice
but no as the mixed state with long-rang phase coherence. Most authors find
this transition \cite{macikeda}. And few authors \cite{moore} only state
that this transition is absent. I agree with few authors \cite{moore}. But
the absence of the solidification transition does not mean that the
transition into the Abrikosov state is absent, because the definition of
the phase coherence by the correlation function is unsuited for the mixed
state. According to the right definition of the phase coherence the
Abrikosov state is the mixed state with long-rang phase coherence but no
the vortex lattice.

	Thus, the Abrikosov solution \cite{abrikos} gives qualitatively
incorrect result because the mean field approximation is not valid for
description of the mixed state of type II superconductors. The
transition in the Abrikosov state in ideal superconductor can be first
order only and can not be second order. The sharp change of the
resistive properties must be observed at this transition. But this sharp
change is observed in few bulk samples with weak disorder \cite{nik81l,
welp} only. The kink in $YBa_{2}Cu_{3}O_{7-x}$ was observed first
in 1990 year \cite{kwok} only when enough quality single crystals were
obtained. No sharp change is observed both in thin films with weak disorder
\cite{nik95l} and in all samples with strong disorder. The transition
into the Abrikosov state in bulk superconductors becomes smooth with
increasing of disorder amount \cite{fendrich}.

	The absence of any features of the resistive properties in amorphous
$Nb_{1-x}O_{x}$ films was interpreted in our paper \cite{nik95l} as the
absence of the transition into the Abrikosov state down to very low fields.
But Theunissen and Kes \cite{kes97} contend that we do not observe any
feature because our measuring current is extremely high in comparison to
the critical current. I agree with the Theunissen and Kes that the vortex
pinning appearance can not be observed if the measuring current is
extremely high in comparison to the critical current. But I contend that
the transition from the paraconducting regime to the vortex flow regime can
not be without features in the ideal sample.

	The fluctuation decreases the vortex flow resistivity as well as the
resistivity above the transition into the Abrikosov state \cite{maki71}.
The fluctuation value increases near the transition. Therefore features
ought be expected at the transition from the paraconducting regime to the
vortex flow regime. Sharp feature is observed in enough homogeneous bulk
superconductors \cite{nik81j}. At the transition into the Abrikosov state
not only the vortex pinning appears but also the vortex flow resistivity
decreases sharply. Below the transition the vortex flow resistivity
dependence has a minimum \cite{nik81j}. Such dependence differ
qualitatively from the mean-field vortex flow resistivity dependence
\cite{kopnin} and is explained in \cite{nik81j} by fluctuation influence.

	The authors of the paper \cite{kes97} state that these features of the
vortex flow dependencies coincide with the peak effect in the critical
current. But this is right no always. Our investigations \cite{nik85} have
shown that these features are observed in all enough homogeneous bulk
samples both with and without the peak effect. And only in no enough
homogeneous samples the "classical" flux flow resistivity dependencies
\cite{kim64} are observed. Therefore these features ought be considered as
universal for homogeneous bulk superconductors.

	The features of the vortex flow resistivity are observed in paper
\cite{kes97} (see the inset of Fig.7). These features are observed below
the field value where the vortex pinning appears. As it was shown above the
vortex pinning is the consequence of the phase coherence. Consequently the
features of the vortex flow resistivity are observed below the phase
coherence appearance in thin films \cite{kes97} as well as in bulk
superconductors \cite{nik81j}. But these features are no sharp in thin
films \cite{kes97}.

	 The comparison of the pinning values (made in \cite{kes97}) shows that
the amount of disorder in our amorphous $Nb_{1-x}O_{x}$ films
\cite{nik95l} is smaller than in the a-NbGe films used in \cite{kes97}.
Therefore the features at the transition from the paraconducting regime to
the vortex flow regime must be more sharp in our films. The features of the
vortex flow resistivity can be observed at high measuring current. But no
features were observed on the resistivity dependencies of the amorphous
$Nb_{1-x}O_{x}$ films in our paper \cite{nik95l}. Therefore I think that
the phase coherence appearance in thin film is not universal, but that it
depends on the amount of disorder: the phase coherence appears in a middle
field in the a-NbGe films with an intermediate strength of disorder
\cite{kes97} and does not appear down to very low fields in the
$Nb_{1-x}O_{x}$ films with extremely small pinning \cite{nik95l}.

	This claim is confirmed also by our theoretical results \cite{nik95b}.
The transition into the Abrikosov state is connected with the Abrikosov
parameter $\beta_{a} = \overline{n_{s}^{2}}/\overline{n_{s}}^{2}$. Where
$\overline{n_{s}^{2}} = (\int_{V}dRn_{s}^{2})/V$ and $\overline{n_{s}} =
(\int_{V}dRn_{s})/V$ is the spatial average value of the superconducting
electron density. Kleiner, Roth and Auther \cite{kleiner} have shown that
the minimum possible value $\beta_{a}$ is equal $\beta_{A} \simeq 1.16$ if
$n_{s}(R)$ are lowest-Landau-level (LLL) functions. The minimum $\beta_{a}$
value only is considered in the mean field approximation. Therefore the
triangular vortex lattice appears below $H_{c2}$ according to this
approximation because this state corresponds the minimum $\beta_{a}$ value
\cite{huebener}. But the exact thermodynamic average $<\beta_{a}>$ is not
equal $\beta_{A}$ in the consequence of the fluctuation. The $<\beta_{a}>$
value changes from 2 above the $H_{c2}$ critical region to a value closed
to $\beta_{A}$ below the $H_{c2}$ critical region \cite{nik95b}. The
$<\beta_{a}>$ dependence of two-dimensional superconductor was calculated
in work \cite{macdonal} for whole LLL region. $<\beta_{a}>$ is not equal
$\beta_{A}$ at a finite temperature. Therefore the transition into the
Abrikosov state can be when the $<\beta_{a}>$ value is smaller then a
critical value, i.e. when $<\beta_{a}> - \beta_{A} < (\beta_{a} -
\beta_{A})_{c}$. According to \cite{nik95b} the $(\beta_{a} -
\beta_{A})_{c}$ value of two-dimensional ideal superconductor decreases
with sample size increasing. In real case the sample size ought to be
replaced by a effective distance between disorders. The $<\beta_{a}> -
\beta_{A}$ value decreases with the magnetic field (or temperature)
decreasing \cite{macdonal,nik95b}. Therefore according to \cite{nik95b}
the position of the transition into the Abrikosov state depends on the
amount of disorder in thin films.

	Thus, the absence of any sharp change of the resistive properties in
thin films with weak disorder \cite{nik95l} can be explained by the absence
of the transition into the Abrikosov state down to very low fields. The
absence of the long-rang phase coherence below $H_{c2}$ in these films is
confirmed by a recent investigation of the nonlocal resistivity
\cite{nik97}. But it is obvious that the long-rang phase coherence is
appeared in superconductors with strong \cite{fendrich} and intermediate
\cite{kes97} strength of disorder, because the pinning effect is observed
in these cases. Consequently, the phase coherence appearance in
superconductors with strong and intermediate strength of disorder differs
from the ideal case. The transition into the Abrikosov state is smooth in
these superconductors. The length of the phase coherence does not change by
jump but increases gradually with the magnetic field (or the temperature)
decreasing.

	In the present work the phase coherence appearance in thin films with
strong disorder is considered. The resistive properties of the thin
$Nb_{1-x}O_{x}$ films with weak and strong disorder is compared. In order
to explain the difference of the phase coherence appearance in
superconductors with strong disorder from the ideal case it is proposed to
return to the Mendelssohn model.

\section{SAMPLE PREPARATION AND CHARACTERISTICS}

	The $Nb_{1-x}O_{x}$ films were produced by magnetron sputtering of Nb in
an atmosphere of argon and oxygen. Changing the oxygen we produced the
films with different oxygen contents. The transmission electron microscopy
investigation shown that the films with small oxygen contents have small
grain structure whereas the films with greater oxygen contents are
amorphous. The temperature of the superconducting transition, $T_{c}$, of
the films decreases with the oxygen content increasing. At the oxygen
content $x > 0.2$ the critical temperature $T_{c} < 2 \ K$. In the present
work the amorphous films with $x \simeq 0.2$ and the films with small
($\simeq 10 \ nm$) grain structure with $x \simeq 0.08$ were used. The
oxygen content was determined by Auger analysis with a relative error 0.3.
The critical temperature of used amorphous films $T_{c} = 1.8 - 3 \ K$ and
of the used films with small grain structure $T_{c} = 5.7 \ K$.
$(dH_{c2}/dT)_{T=Tc} = 22 \ kOe/K$ for the amorphous films and
$(dH_{c2}/dT)_{T=Tc} = 6 \ kOe/K $ for the crystalline films. The
temperature dependence of normal resistivity of both films is weak. The
normal resistivity $\rho_{n} = 3 \ 10^{-7} \Omega m$ of the films with
small grain structure decreasing weakly with temperature decreasing.
Whereas the resistivity $\rho_{n} = 20 \ 10^{-7} \Omega m$ of the amorphous
films increases with decreasing temperature. This change can be connected
with weak localization.

	A perpendicular magnetic field up to 50 kOe produced by a
superconducting solenoid was measured with relative error 0.0005. The
temperature was measured with a relative error 0.001. The resistivity was
measured with a relative error 0.0001.

\section{RESULTS AND DISCUSSION}

	The scaling of the resistivity was observed in amorphous $Nb_{1-x}O_{x}$
\cite{nik95l} and a-NbGe \cite{kes97} films with weak disorder. This
scaling is general consequence of the fluctuation Ginzburg-Landau theory in
the lowest Landau level (LLL) approximation \cite{nik96}. The scaling of
various properties is consequence of the scaling of the spatial average
value of the superconducting electron density $\overline{n_{s}}$
\cite{nik96}. Strictly speaking the scaling of $\overline{n_{s}}$ is
valid in the state without the phase coherence only because the
$\overline{n_{s}}$ value changes at the phase coherence appearance. But
this change is very small \cite{shilling}. Therefore the dependencies of
the specific heat \cite{ferrant}, magnetization and other thermodynamic
properties depended only on the $\overline{n_{s}}$ value are close to the
scaling in the states both with and without the phase coherence.

	But the transport properties depend strongly on the phase coherence.
Therefore the resistivity dependencies can be close to the scaling in the
state without the phase coherence only. The sharp deviation of the
conductivity dependencies from the scaling is observed in bulk
superconductors at the transition into the Abrikosov state
\cite{nik83,nik85}. Consequently, the scaling of the resistivity is
evidence of the absence of the phase coherence and the absence of this
scaling is evidence of the phase coherence. Therefore we can use the
comparison of the resistive dependencies with the scaling for the
investigation of the phase coherence appearance.

	In contrast of the results \cite{nik95l,kes97} the resistivity
dependencies of films with strong disorder clearly deviate from the scaling
Fig.1. According to the scaling law the $\Delta \sigma (Gi_{2D}ht)^{0.5}/t$
is a universal function of $(t-t_{c2})/(Gi_{2D}ht)^{0.5}$. Here $\Delta
\sigma = \sigma(t,h) - \sigma_{n}$ is the excess conductivity of
two-dimensional superconductor (i.e. thin film); $\sigma_{n} = 1/\rho_{n}$
is the normal conductivity of the film; $Gi_{2D} = k_{B}T_{c}/H_{c}^{2}(0)d
\xi^{2}(0)$ is the Ginzburg number of two-dimensional superconductor; $t =
T/T_{c}$; $h = H/H_{c2}(0)$; $H_{c2}(0) = -T_{c}(dH_{c2}/dT)_{T=Tc}$;
$H_{c}(0)$ is the thermodynamic critical field at T = 0; $\xi(0)$ is the
coherence length at T = 0; d is the film thickness. In Fig.1 the
$(1+(\Delta \sigma/\sigma_{n})(h/t)^{0.5})^ {-1}$ versus
$(t-t_{c2})/(ht)^{0.5}$ dependencies of the $Nb_{1-x}O_{x}$ film with small
grain structure in different magnetic fields are shown. (The same
coordinates were used on Fig.4 of Ref.\cite{nik95l}). The Fig.1
demonstrates clearly that the scaling is not observed in the film with
strong disorder.

	The excess conductivity of the film with strong disorder deviates
clearly from the scaling already above $H_{c2}$. As demonstrated in Fig.2
the $\Delta \sigma (Gi_{2D}ht)^{0.5}d/t\sigma_{0}$ versus
$(h-h_{c2})/(Gi_{2D}ht) ^{0.5}$ dependencies of the $Nb_{1-x}O_{x}$ film
with small grain structure begins to deviate from the one of the amorphous
$Nb_{1-x}O_{x}$ film at $h \simeq 1.05h_{c2}$. Where $\sigma_{0} =
e^{2}/\hbar$; $h_{c2} = H_{c2}/H_{c2}(0)$. The $h_{c2}$ value is
determined by the comparison of the experimental and theoretical
paraconductivity dependencies in the linear approximation region. The
reliability of this method has been demonstrated at the investigation of
the paraconductivity in bulk superconductors. The distinction of the
position of the transition into the Abrikosov state from $H_{c2}$ has
been discovered first by the using of this method \cite{nik81j}. Later
\cite{nik84} this result was confirmed by determination of $H_{c2}$ value
from magnetization measurement.

	The deviation of the conductivity dependence from the scaling in thin
films with strong disorder differs from the one in bulk superconductor with
weak disorder. The sharp deviation is observed in bulk superconductors and
it coincides with the qualitative change of the current-voltage
characteristic form \cite{nik83,nik85}. The current-voltage characteristics
become non-Ohmic. The deviation in the thin film is smooth and is observed
above the magnetic field value where the non-Ohmic current-voltage
characteristics are observed. For example at T = 4.2 K the non-Ohmic
current-voltage characteristics are observed in the $Nb_{1-x}O_{x}$ film
with small grain structure at $H/H_{c2} \simeq 0.7$ ($H_{c2} = 9.0
kOe$) whereas the deviation is observed at $H/H_{c2} \simeq 1.05$ (Fig.2).

	At $0.2 < H/H_{c2} < 0.65$ the current-voltage characteristics can be
described by the relation

$$E=E_{0} \sinh(j/j_{0}) \eqno{(3)}$$

(see Fig.3). According to this relation at $j \ll j_{0}$, $E \simeq
E_{0}j/j_{0} = \rho_{TAFF}j$. Where $\rho_{TAFF} = E_{0}/j_{0}$ is a
thermally activated linear resistivity \cite{brandt95}. The $\rho_{TAFF}$
value decreases strongly with magnetic field decreasing: from $3 \ 10^{-9}
\Omega m$ at H = 6 kOe to $10^{-15} \Omega m$ at H = 2 kOe (Fig.3).
Expected $\rho_{TAFF}$ values at $H/H_{c2} < 0.2$ and at $H/H_{c2} >
0.65$ can be estimated by extrapolation of the $E_{O}(H)$ and
$j_{0}(H)$ dependencies (see Fig.5). According to this extrapolation
$\rho_{TAFF} = 1.2 \ 10^{-8} \Omega m$ at H = 7 kOe is close to
resistivity value $1.4 \ 10^{-8} \Omega m$ measured at H = 7 kOe whereas
$\rho_{TAFF} \simeq 10^{-20} \Omega m$ at H = 1 kOe is too little for
experimental measurement. Therefore I may assume that the relation (3) is
valid in more wide region than $0.2 < H/H_{c2} < 0.65$. But it is obvious
that the relation (3) can be valid only if $\rho_{TAFF} \ll \rho_{scal}$.
Where $\rho_{scal}$ is a resistivity value according to the scaling law
(i.e. the resistivity in films with weak disorder \cite{nik95l}). The
relation $\rho_{TAFF}/\rho_{scal}$ at T = 4.2 K ($H_{c2} = 9.0 \ kOe$) is
equal 0.38; 0.07; 0.02; 0.004; 0.0008; 0.00001 at H = 8; 7; 6; 5; 4; 2 kOe.
I use the measured linear resistivity at H = 8 and 7 kOe as
the $\rho_{TAFF}$ value. Consequently the relation (3) can not be valid
near $H_{c2} = 9.0 \ kOe$ where $\rho_{TAFF} \simeq \rho_{scal}$.

	The smooth deviation of the conductivity dependence from the scaling
Figs.1,2 means that the length of the phase coherence increases smoothly.
This result differs strongly from the ideal case. As it was shown above the
ideal phase coherence appearance must be sharp transition because the
fluctuation is zero-dimensional in thin film without disorder placed in
high perpendicular magnetic field. The observed difference of the phase
coherence appearance from the ideal case can be explained by increasing of
the effective dimensionality of the fluctuation (from zero to one in film)
in samples with strong disorder.

	The limit case of strong disorder is the Mendelssohn's "sponge"
\cite{mendelss}. The Mendelssohn's "sponge" is a model proposed more than
sixty years ago for explanation of the magnetic properties of some
superconductors (now we call these superconductors as type II
superconductors). Mendelssohn has assumed that the magnetic field can
penetrate inside superconducting region because this region is
superconducting sponge. Indeed, the magnetic and resistive properties the
superconducting sponge are rather like the one of type II superconductors
with strong disorder. The magnetization dependence is irreversible in both
cases. The critical current can be great in high magnetic field.

	The Abrikosov vortex can exist in the superconducting sponge with the
phase coherence as well as in type II superconductor. The phase coherence
in the superconducting sponge can be defined by the relation (2). The
Abrikosov vortex is defined as a singularity of the wave function. The
integral $\int_{l}dR d\phi/dR$ around this singularity is equal $2\pi$. Let
us consider the two-dimensional superconducting sponge with long-rang phase
coherence placed in a magnetic field (see Fig.4A). The integral
$\int_{l}dR d\phi/dR$ around any cell of the sponge must be equal $2\pi
n$. Where n is an integer. If $H/\Phi_{0}$ is smaller than a cell density
then n = 1 or 0 in a state with minimum value of the free energy. The free
energy value increases at $n > 1$ because the superconducting current value
increases in this case in accordance with the relation (2). We can say that
a cell contains the Abrikosov vortex if n = 1 and the Abrikosov vortex is
absent if n = 0 (see Fig.4). The free energy has minimum in the
state with homogeneous vortex density (Fig.4).

	The Abrikosov state in thin film is similar to the two-dimensional
Mendelssohn sponge Fig.4. The main difference is that the Abrikosov vortex
destroys superconductivity near itself in the Abrikosov model
(Fig.4B) whereas in the Mendelssohn model it occupies a nonsuperconducting
region (Fig.4A). According to the relation (2), near a vortex
$\lambda_{L}^{2}j_{s} = 2mcv_{s}/e \simeq \Phi_{0}/2\pi r - Hr/2$. Here r
is the distance from the vortex center; $v_{s}$ is the velocity of the
superconducting electrons. I consider the case $\lambda_{L} \gg \xi > d$.
$B(r) \simeq H$ in this case. At a $r_{c}$ value, $v_{s}$ is equal the
critical velocity \cite{tinkha75}. Therefore $n_{s} = 0$ at $r < r_{c}$.
The calculations show that $r_{c} \simeq \xi$. Disordes acting as pinning
centers have lower superconducting parameters. In the limit case they are
nonsuperconducting inclusions (Fig.4B). The Mendelssohn sponge (Fig.4A) is
a superconductor with nonsuperconducting inclusions size of which b larger
than the correlation length, $b > \xi$. Thus, we may consider real
superconductors with disorder as intermediate cases between the
Mendelssohn's \cite{mendelss} and Abrikosov's \cite{abrikos} models.

	Because the Abrikosov vortex destroys superconductivity near itself at
$r < r_{c} \simeq \xi(T)$ and the distance between the Abrikosov vortexes
is equal approximately $(\Phi_{0}/H)^{0.5}$ we may consider the thin film
with strong disorder as two-dimensional Mendelssohn sponge with variable
width of superconducting threads $w(T,H) \simeq (\Phi_{0}/H)^{0.5} -
\xi(T)$. The w(T,H) value increases with the T and H decreasing. The above
relation can be valid at enough great w value only. Therefore this relation
can not be used for the evaluation a critical field $H_{c,sp}$ at which w =
0. One ought expected that $H_{c,sp} \simeq H_{c2}$ in film with strong
disorder. The mean field critical field of the superconducting sponge with
cell size $ \gg \xi$ and $w < \xi$ can be appreciably higher than $H_{c2}$.
This critical field, as well as the critical field of thin film in parallel
magnetic field \cite{tinkha75}, is equal $H_{c,sp} = 3^{0.5}\Phi_{0}/\pi
\xi w = 3^{0.5}2 H_{c2}\xi/w$.

	I can not describe qualitatively the resistive dependence of the
Mendelssohn sponge with variable w(T,H) value. But it is obvious that this
dependence is smooth because the Mendelssohn sponge is a one-dimensional
system. In one-dimensional superconductor \cite{grunberg} the length of the
phase coherence increases smoothly with temperature decreasing below
$T_{c}$. In consequence of this the resistive dependence is smooth also.
For example see the resistive transition of bulk superconductor in parallel
magnetic field in the paper \cite{nik93}. It is obvious that the
$\rho_{TAFF}/\rho_{scal}$ value decreases in the consequence of the
increasing of the length of the phase coherence. When the length of the
phase coherence has increased up to sample size the $\rho_{TAFF}$ value
becomes $\ll \rho_{scal}$. Consequently the length of the phase coherence
of the $Nb_{1-x}O_{x}$ film with small grain structure  at T = 4.2 K
increases from $\simeq (\Phi_{0}/H)^{0.5}$ at $H = 1.05H_{c2} = 9.5 kOe$
(Fig.2) to sample size at H = 6 - 7 kOe (see above).

	A crossover to the vortex creep regime takes place in low magnetic field
where $\rho_{TAFF} \ll \rho_{scal}$ (Fig.3). Below I compare the
experimental data (Fig.3) with theoretical results obtained in the
Kim-Anderson model \cite{anders64} and obtained in a model of the vortex
creep in the Mendelssohn sponge. The Kim-Anderson model \cite{anders64}
describes the vortex creep in type II superconductors with pinning.
According to \cite{anders64} the current-voltage characteristics in the
vortex creep regime are described by the relation (3) in which $j_{0} =
k_{B}T/BV_{j}l_{j}$ (see \cite{brandt95}). Here $V_{j}$ is jumping volume
and $l_{j}$ is the jump width. Because the film is thin $V_{j} = dS_{j}$,
where $S_{j}$ is the jumping area. We can estimate the $S_{j}l_{j} =
k_{B}T/j_{0}Hd$ value from the experimental dependencies shown partly on
Fig.3. The $(S_{j}l_{j})^{1/3}$ value is plotted versus the
$(\Phi_{0}/H)^{1/2}$ value in Fig.5A. The distance between the Abrikosov
vortex in the triangular lattice $(2\Phi_{0}/3^{1/2}H)^{1/2}$ is shown also
in Fig.5A. According to Fig.5A the $(S_{j}l_{j})^{1/3}$ value does not
exceed $(2\Phi_{0}/3^{1/2}H)^{1/2}$ value. Therefore I can conclude that
the individual vortex creep takes place in our film.

	The vortex creep can be in the Mendelssohn sponge if the vortex can
jump from one to the other cell (see Fig.4A). A vortex can jump to next cell
of the Mendelssohn sponge if superconductivity has been destroyed in the
cell wall. The energy must increase at this on $\xi(T) dwf_{GL}$, where
$f_{GL}$ is the difference of the free-energy density in the normal and the
superconducting phase \cite{tinkha75}. Consequently the probability of the
jump is proportional to $\exp(\xi(T) dwf_{GL}/k_{B}T)$. According to the
Ginzburg-Landau theory $f_{GL} = (H_{c}^{2}(T)/8\pi)(1 - (\hbar mv_{s}\xi
)^{2}) \cite{tinkha75}$. Because the magnetic flux inside the cell is
smaller than $\Phi_{0}$ a superconducting current is around the cell (see
relation (2)). Therefore the transport current increases the velocity of
the superconducting electrons in one side of the cell and decreases in
opposite side. Consequently the transport current increases the rate of
thermally activated jumps of vortexes in one side and decreases it in
opposite side (Fig.4A). The transport current change weakly the $v_{s}$
value. Therefore we can use following relation for the vortex velocity

$$v_{vor} = 2l_{j}\omega_{0} \exp(\frac{\xi dwf_{GL}}{k_{B}T})
\sinh(\frac{\xi dwv_{tr}}{k_{B}T}\frac{df_{GL}}{dv_{s}}) \eqno{(4)}$$

where $\omega_{0} $ is an attempt frequency; $v_{tr}$ is a change of the
$v_{s}$ value in consequence of the transport current. Because the voltage
is proportional the vortex velocity the current-voltage characteristics of
the Mendelssohn sponge in the creep regime conforms to the relation (3).
$\xi wdv_{tr}(df_{GL}/dv_{s}) = (\xi d\Phi_{0}/8\pi^{2})(1 -
(\xi/b)^{2}(1 - HS_{cell}/\Phi_{0})^{2})(1 - HS_{cell}/\Phi_{0})j
\simeq (\xi d\Phi_{0}/8\pi^{2})j$, where  $S_{cell}$ is cell area; j is the
transport current density. Consequently, the $j_{0}$ value (see the relation
(3)) of the two-dimensional Mendelssohn sponge is equal $j_{0} \simeq
8\pi^{2}k_{B}T/\xi d\Phi_{0}$. For the parameter values of our film ($\xi
\simeq 20 nm$, d = 20 nm), $j_{0} \simeq 5 \ 10^{9} \
A/m^{2}$ at T = 4.2 K. The experimental $j_{0}$ values (Fig.3) are smaller
approximately by a factor of 50 - 100.

	The observed decreasing of the $E_{0} = Hl_{j}\omega_{0} \exp(\xi
dwf_{GL}/k_{B}T)$ value with magnetic field decreasing (see Fig.3) can be
explained qualitatively by the increasing of the w value. The experimental
dependence $\ln(E_{0}/E_{0,Hc2})$ versus $(H_{c2}/H)^{0.5} - 1$ is plotted
in Fig.5B. Where $E_{0,Hc2}$ is a normalizing voltage value. I following a
model according to which $w(H) \simeq \xi ((H_{c2}/H)^{0.5} - 1)$. The
dependence shown on Fig.5B corresponds qualitatively to this model.
$E_{0,Hc2} = 2Hl_{j}\omega_{0}$ in this case. The slope of the dependence
$\ln(E_{0}/E_{0,Hc2})$ versus $(H_{c2}/H)^{0.5} - 1$ is equal $-\xi^{2}
df_{GL}/k_{B}T$ in this model. According to Fig.5B $\xi^{2} df_{GL}/k_{B}T
\simeq 16$ whereas the $\xi^{2}dH_{c}^{2}/8\pi $ value of our
$Nb_{1-x}O_{x}$ film at T = 4.2 K is equal approximately $100k_{B}T$. I can
conclude that the vortex pinning in the $Nb_{1-x}O_{x}$ film with small
grain structure is close to the limit case of strong disorder but our
simple Mendelssohn model describes the vortex creep regime qualitatively
only.

\section{CONCLUSIONS}

	The smooth phase coherence appearance in thin films with strong disorder
can be explained qualitatively by the increasing of the effective
dimensionality of the fluctuation. According to this explanation
the smooth transition into the Abrikosov state observed in majority of
samples differs qualitatively from the ideal case. Thin films are
considered in the present work because the case of bulk superconductors is
more difficult. But I think that the transition into the Abrikosov state
in bulk superconductors is smoothed out by disorders in consequence of
similar cause as considered in this paper. Real superconductors with
disorder can be considered as an intermediate case between the Mendelssohn
and Abrikosov models. Because intermediate cases are difficult for
theoretical description the resistive properties even thin films with
disorder can be described qualitatively only.

	The second critical field $H_{c2}$ is no critical point not only in
superconductors with weak disorder but also in superconductors with strong
disorder. The phase coherence appears below $H_{c2}$ in superconductors
with weak disorder \cite{nik81l,welp} and above $H_{c2}$ in superconductors
with strong disorder.

	The sharp transition into the Abrikosov state predicted by the
fluctuation theory in ideal case is observed in bulk superconductors with
weak disorder \cite{nik81l,welp} only. No sharp transition is observed in
thin films with weak disorder \cite{nik95l}. This difference can be
explained by difference of the fluctuation value in three- and
two-dimensional superconductors.

	The mean field approximation can not be used for the description of the
mixed state in the thermodynamic limit, because according to \cite{maki71}
the fluctuation value in the Abrikosov state calculated in the linear
approximation increases with superconductor size increasing. But according
to \cite{maki71} the mean field approximation can be valid in
superconductors with finite sizes: at $h_{c2} - h \gg
(\ln(L/\xi))^{2/3}Gi_{3D}^{1/3}(th)^{2/3}/0.16$ in bulk (three-dimensional)
superconductor and at $h_{c2} - h \gg \ (L/0.06\xi)(Gi_{2D}th)^{1/2}$ in
thin film (two-dimensional superconductor). Here L is sample size across a
magnetic field; $\xi$ is coherence length; $Gi_{3D} =
(k_{B}T_{c}/H_{c}^{2}(0)\xi^{3}(0))^{2}$ is the Ginzburg number of
three-dimensional superconductor. For real values $L = 1 mm$ and $\xi =
10^{-5} mm$, $\ln(L/\xi) \simeq 10$ whereas $L/\xi = 10^{5}$. Because for
conventional superconductors $Gi_{3D} = 10^{-5} - 10^{-11}$ and $Gi_{2D} =
10^{-2} - 10^{-6}$ the mean field approximation is not valid in a narrow
region near $H_{c2}$ only in bulk superconductors and in whole mixed state
of thin films. Therefore the transition into the Abrikosov state in bulk
superconductors can be like the ideal transition. Whereas in thin films the
phase coherence can appear in consequence of disorder only.

\section{ACKNOWLEDGMENT}

	I thank the International Association for the Promotion of Co-operation
with Scientists from the New Independent States (Project INTAS-96-0452) and
the National Scientific Council on "Superconductivity" of SSTP "ADPCM"
(Project 98013) for financial support.

\newpage

Figure Captions

Fig.1. The $(1+(\Delta \sigma/\sigma_{n})(h/t)^{0.5})^ {-1}$ versus
$(t-t_{c2})/(ht)^{0.5}$ dependencies of the $Nb_{1-x}O_{x}$ film with small
grain structure in magnetic fields H = 4 kOe (h = 0.12) and H = 12 kOe (h =
0.36). Film thickness d = 20 nm. The measuring current $j = 10^{6} \
A/m^{2}$. The current-voltage characteristics are Ohmic in the shown region
of temperature and magnetic field values.

Fig.2. The $\Delta \sigma (Gi_{2D}ht)^{0.5}d/t\sigma_{0}$ versus
$(h-h_{c2})/(Gi_{2D}ht) ^{0.5}$ dependencies of the $Nb_{1-x}O_{x}$ film
with small grain structure ($T_{c} = 5.7 \ K$; $Gi_{2D} \simeq 0.0005$) at
T = 4.2 K (curve 1) and the amorphous $Nb_{1-x}O_{x}$ film ($T_{c} = 2.37 \
K$; $Gi_{2D} \simeq 0.00015$) at T = 1.72 K (curve 2). $t = T/T_{c} \simeq
0.7$ for both curve. Thickness of both films is equal d = 20 nm.

Fig.3. Current-voltage characteristics of the $Nb_{1-x}O_{x}$ film with
small grain structure (with d = 20 nm) in magnetic fields H =
6 kOe (1); H = 5 kOe (2) and H = 4 kOe (3) at T = 4.2 K ($H_{c2} = 9.0 \
kOe$). The lines denote the $E = E_{0}sinh(j/j_{0})$ dependencies with
$E_{0} = 0.27 \ V/m$ and $j_{0} = 0.90 \ 10^{8} \ A/m^{2}$ for H = 6 kOe;
$E_{0} = 0.024 \ V/m$ and $j_{0} = 0.60 \ 10^{8} \ A/m^{2}$ for H = 5 kOe;
$E_{0} = 0.0015 \ V/m$ and $j_{0} = 0.43 \ 10^{8} \ A/m^{2}$ for H = 4 kOe.

Fig.4. Sketches of the two-dimensional Mendelssohn sponge (A) and of
two-dimensional superconductors with disorders (pinning centers) (B) in the
Abrikosov state. Superconducting regions are dark. Nonsuperconducting
regions are light. Pinning centers are represented by rectangles. Cells
with the Abrikosov vortex are marked by 1, without the vortex - 0. Size of
the nonsuperconducting regions b in Mendelssohn sponge is larger than the
correlation length (A) whereas the one of pinning centers is smaller than
the correlation length (B). The Abrikosov vortexes destroy of
superconductivity in circles with radius $r_{c} > b$ (B). Transport current
j increases the rate of thermally activated jumps of vortexes to the right
and decreases it to the left (A). w is width of superconducting threads
(A).

Fig.5. A) The $(S_{j}l_{j})^{1/3} = (k_{B}T/j_{0}Hd)^{1/3}$ versus
$(\Phi_{0}/H)^{1/2}$ dependence (points with error bars). The line denotes
the distance between the Abrikosov vortexes in the triangular lattice. B)
The $\ln(E_{0}/E_{0,Hc2})$ versus $(H_{c2}/H)^{0.5} - 1$ dependence.
$E_{0,Hc2} = 6.9 \ V/m$; $H_{c2} = 9.0 \ kOe$. The slope of the line is
equal -16. Points with error bars in A) and B) are data obtained from the
current-voltage characterictics of the $Nb_{1-x}O_{x}$ film with small
grain structure (d = 20 nm) at T = 4.2 K (see Fig.3).


\begin{thebibliography}{9}

\bibitem{blatter} G.Blatter, M.V.Feigel'man, V.B.Geshkenbein, A.I.Larkin,
and V.M.Vinokur, Rev.Mod.Phys. {\bf66}, 1125 (1994).

\bibitem{nik96} A.V.Nikulov, in {\em {Fluctuation Phenomena in High
Temperature Superconductors}} (Ed. M.Aussloos and A.A.Varlamov) (Kluwer
Academic Publishers, Dordrecht/Boston/London, 1997) p.271.

\bibitem{nik81l} V.A.Marchenko and A.V.Nikulov, Pisma
Zh.Eksp.Teor.Fiz. {\bf34}, 19 (1981) (JETP Lett. {\bf34}, 17 (1981)).

\bibitem{welp} H.Safar, P.L.Gammel, D.A.Huse, D.J.Bishop, J.P.Rice, and
D.M.Ginzberg, Phys.Rev.Lett. {\bf69}, 824 (1992); W.K.Kwok, S.Fleshler,
U.Welp, V.M.Vinokur, J.Downey, G.W.Crabtree, and M.M.Miller, Phys.Rev.Lett.
{\bf69}, 3370 (1992); W.Jiang, N.-C.Yeh, D.S.Reed, U.Kriplani, and
F.Holtzberg, Phys.Rev.Lett. {\bf74}, 1438 (1995).

\bibitem{nik97} A.V.Nikulov, S.V.Dubonos, and Y.I.Koval, J.Low Temp.Phys.
{\bf109}, 643 (1997)

\bibitem{abrikos} A.A.Abrikosov, Zh.Eksp.Teor.Fiz. {\bf32}, 1442 (1957)
(Sov.Phys.-JETP {\bf5}, 1174  (1957) ).

\bibitem{degennes} P.G.De Gennes, {\em {Superconductivity of Metals and
Alloys}} (New York: Benjamin, 1966).

\bibitem{lee72} P.A.Lee and S.R.Shenoy, Phys.Rev.Lett. {\bf 28}, 1025
(1972).

\bibitem{tinkha75} M.Tinkham, {\em {Introduction to Superconductivity}}
(McGraw-Hill Book Company, 1975)

\bibitem{huebener} R.P.Huebener, {\em {Magnetic Flux Structures in
Superconductors}} (Springer-Verlag, Berlin Heidelberg New York, 1919).

\bibitem{london35} F.London and H.London, Proc.Roy.Soc. (London) A
{\bf149}, 71 (1935).

\bibitem{meissner} W.Meissner and R.Ochsenfeld, Naturwiss. {\bf21}, 787
(1933).

\bibitem{campbell} A.M.Campbell and J.E.Evetts, {\em {Critical Currents in
Superconductors}} (Taylor and Francis LTD, London, 1972)

\bibitem{nik90} A.V.Nikulov, Supercond.Sci.Technol. {\bf3}, 377 (1990).

\bibitem{bishop} D.Bishop, Nature {\bf382}, 760 (1996).

\bibitem{shilling} A.Schilling et al., Nature {\bf 382}, 791 (1996).

\bibitem{troible} D.Cribier, B.Jacrot, L.M.Rao, and B.Farnoux,
Phys.Lett. {\bf9}, 106 (1964); U.Essmann and H.Trauble, Phys.Let. A
{\bf24}, 526 (1967).

\bibitem{larkin70} A.I.Larkin, Zh.Eksp.Teor.Fiz. {\bf58},
1466 (1970) (Sov.Phys.-JETP {\bf31}, 784 (1970)).

\bibitem{tesan91} Z.Tesanovic, Phys.Rev. B {\bf44}, 12635 (1991).

\bibitem{macikeda} Z.Tesanovic and L.Xing, Phys.Rev.Lett. {\bf67}, 2729
(1991); Y.Kato and N.Nagaosa, Phys.Rev. B {\bf47}, 2932 (1993); Phys.Rev. B
{\bf48}, 7383 (1993); J.Hu and A.H.MacDonald, Phys.Rev.Lett {\bf 71}, 432
(1993); R.Sasik and D.Stroud, Phys.Rev.Lett {\bf 72}, 2462 (1994);
Phys.Rev.Lett {\bf 75}, 2582 (1995); Phys.Rev. B {\bf 48}, 9938 (1993);
Phys.Rev. B {\bf 49}, 16074 (1994); Phys.Rev. B {\bf 52}, 3696 (1995).

\bibitem{moore} M.A.Moore, Phys.Rev. B {\bf 45}, 7336 (1992); N.Wilkin
and M.A.Moore, Phys.Rev. B {\bf 48}, 3464 (1993); J.A.O'Neill and
M.A.Moore, Phys.Rev.Lett. {\bf 69}, 2582 (1992); J.A.O'Neill and M.A.Moore,
Phys.Rev. B {\bf 48}, 374 (1993); H.H.Lee and M.A.Moore, Phys.Rev. B {\bf
49}, 9240 (1994).

\bibitem{ikeda} R.Ikeda, J.Phys.Soc.Jpn. 65, 3998 (1996); R.Sasik,
D.Stroud and Z.Tesanovic, Phys.Rev. B 51, 3041 (1995).

\bibitem{kwok} W.K.Kwok, U.Welp, G.W.Crabtree, K.G.Vandervoort,
R.Hulscher, and J.Z.Liu, Phys.Rev.Lett. {\bf64}, 966 (1990).

\bibitem{nik95l} A.V.Nikulov, D.Yu.Remisov, and V.A.Oboznov,
Phys.Rev.Lett. {\bf75}, 2586 (1995).

\bibitem{fendrich} J.A.Fendrich et al., Phys.Rev.Lett. {\bf74}, 1210
(1995).

\bibitem{kes97} M.H.Theunissen and P.H.Kes, Phys.Rev.B {\bf55}, 15183
(1997).

\bibitem{maki71} K.Maki and H.Takayama, Prog.Theor.Phys. {\bf46}, 1651
(1971)

\bibitem{nik81j} V.A.Marchenko and A.V.Nikulov, Zh.Eksp.Teor.Fiz.
{\bf80}, 745 (1981) (Sov.Phys.-JETP {\bf53}, 377 (1981)).

\bibitem{kopnin} L.P.Gor'kov and N.B.Kopnin, Usp.Fiz.Nauk {\bf116},
413 (1975) (Sov.Phys. - Uspechi {\bf18}, 496 (1976)).

\bibitem{nik85} A.V.Nikulov, Thesis, Institute of Solid State Physics,
Chernogolovka, 1985.

\bibitem{kim64} Y.B.Kim, C.F.Hempsted, A.R.Strnad, Phys.Rev. {\bf131},
2486 (1963); Phys.Rev. {\bf139}, A1163 (1965).

\bibitem{nik95b} A.V.Nikulov, Phys.Rev. B {\bf52}, 10429 (1995) .

\bibitem{kleiner} W.H.Kleiner, L.M.Roth, and S.H.Autler, Phys.Rev. A
{\bf133}, 1226 (1964).

\bibitem{macdonal} J.Hu and A.H.MacDonald, Phys.Rev. B {\bf 52}, 1286
(1995)

\bibitem{ferrant} R.F.Hassing, R.R.Hake, and L.J.Barnes, Phys.  Rev.Lett.
{\bf30}, 6 (1973);  Farrant, S.P. and Gough C.E., Phys.Rev.Lett. {\bf34},
943 (1975).

\bibitem{nik83} V.A.Marchenko and A.V.Nikulov, Fiz.Nizk.Temp. {\bf9},
816 (1983).

\bibitem{nik84} V.A.Marchenko and A.V.Nikulov, Zh.Eksp.Teor.Fiz. {\bf86},
1395 (1984) (Sov.Phys.-JETP {\bf59}, 815 (1984)).

\bibitem{brandt95} E.H.Brandt, Rep.Progr.Phys. (1995)

\bibitem{mendelss} K.Mendelssohn, Proc.Roy.Soc. {\bf152A}, 34 (1935).

\bibitem{grunberg} L.W.Grunberg and L.Gunther, Phys.Lett. A {\bf38},
463 (1972).

\bibitem{nik93} V.A.Marchenko and A.V.Nikulov, Physica C {\bf210}, 466
(1993).

\bibitem{anders64} P.W.Anderson and Y.B.Kim, Rev.Mod.Phys. {\bf 36}, 39
(1964)

\end{thebibliography}
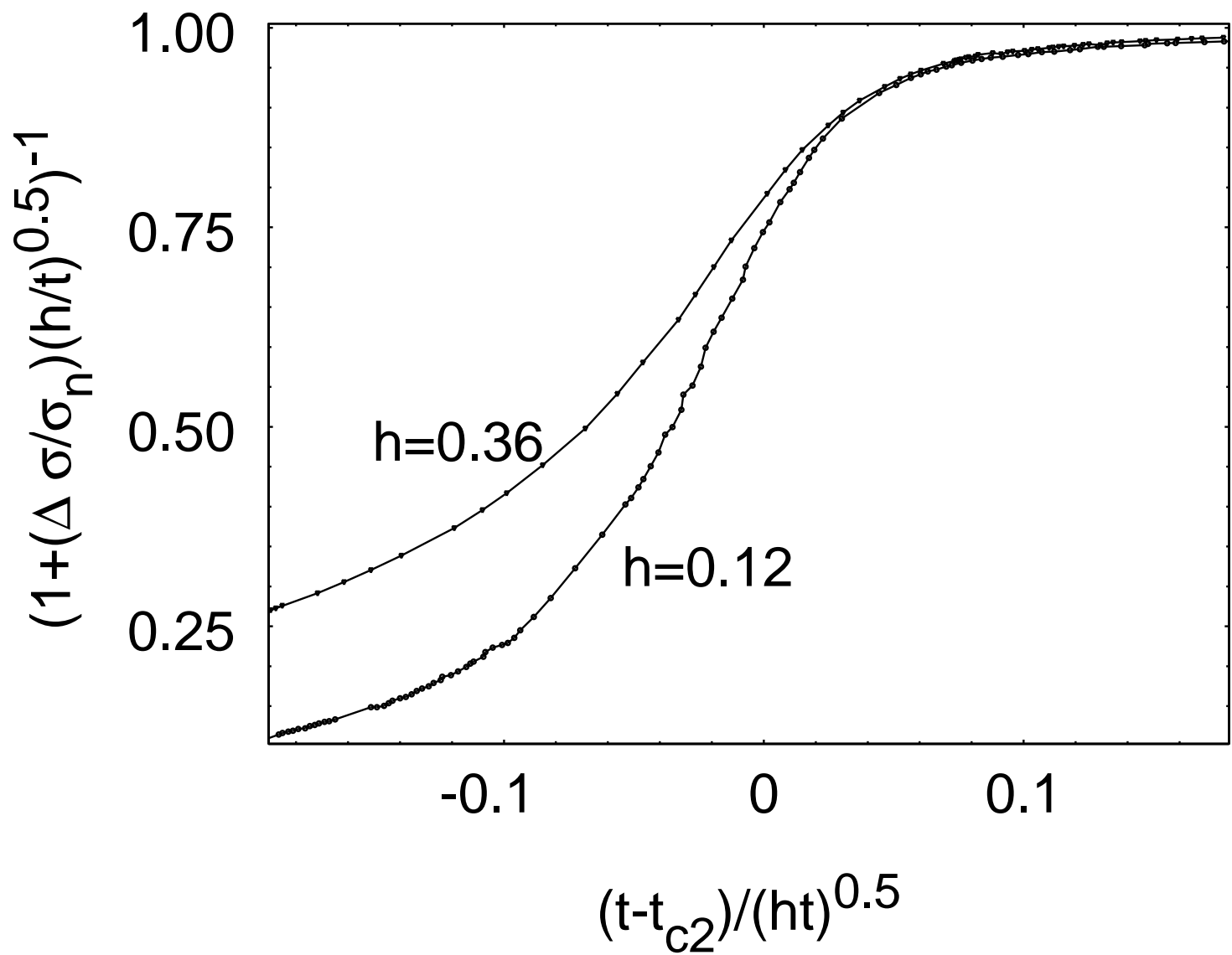
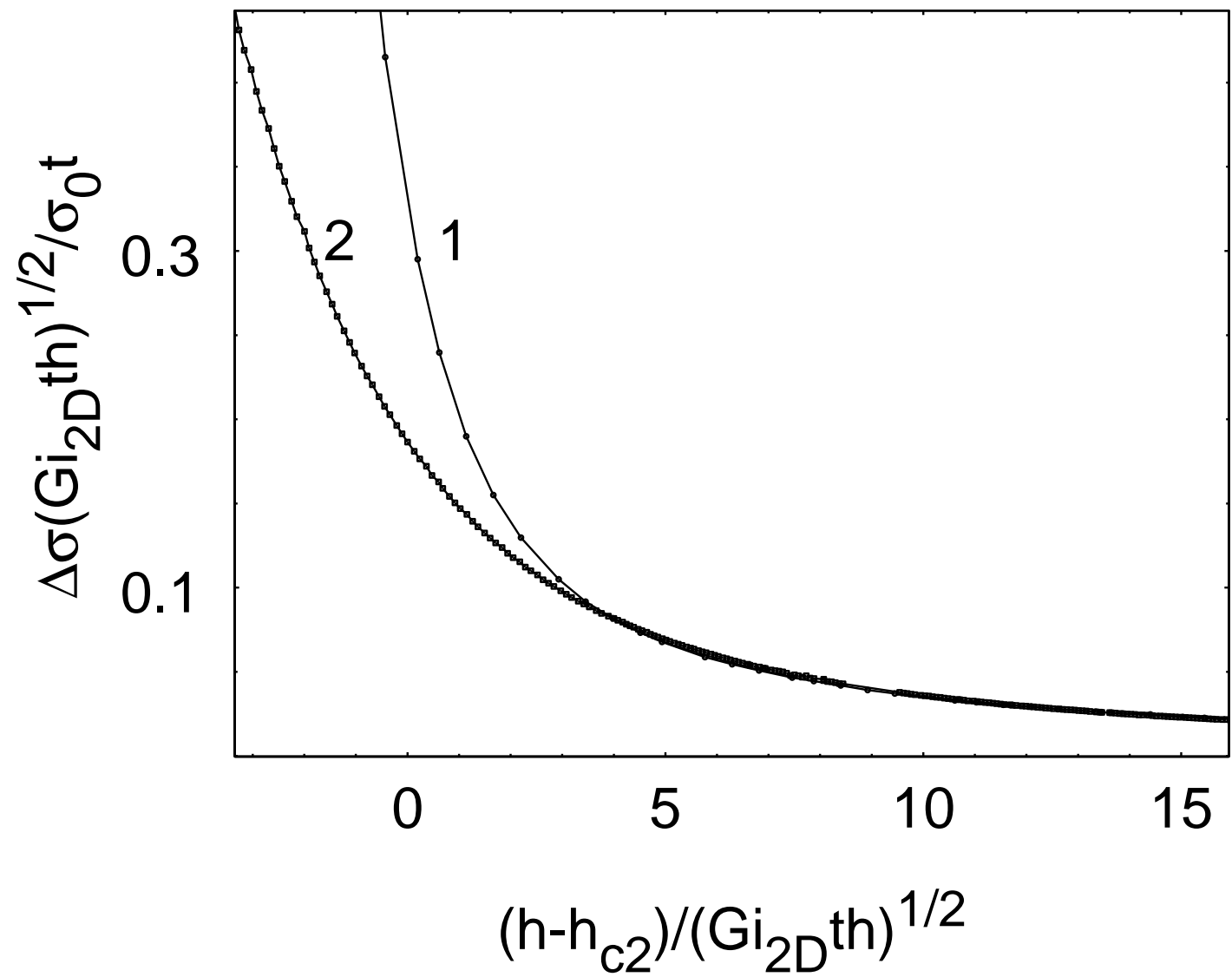
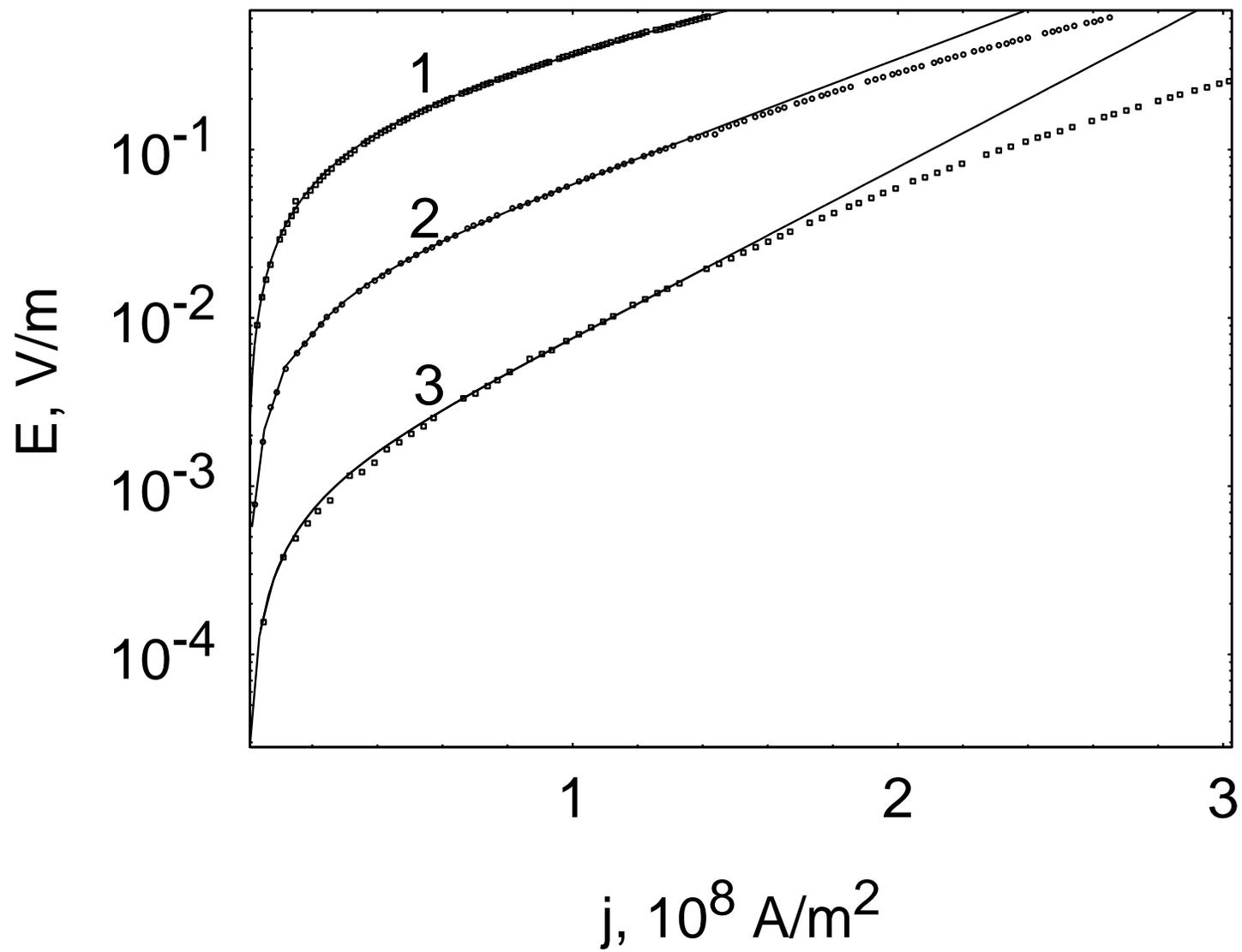
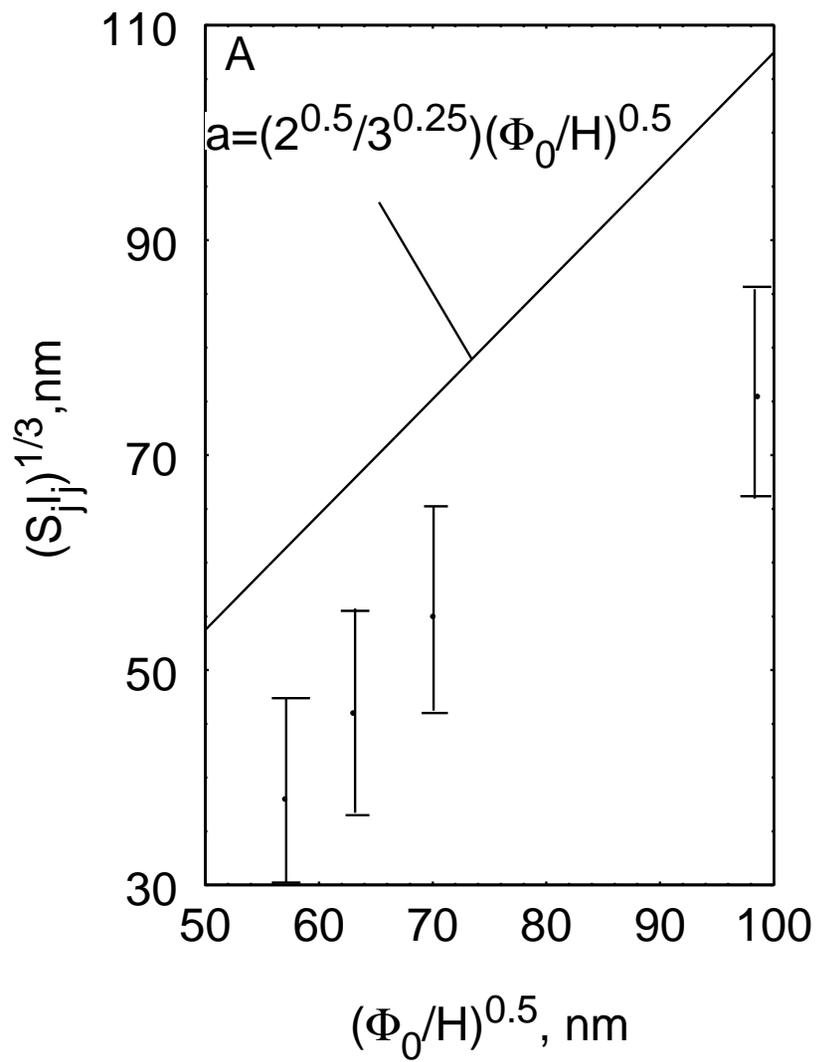
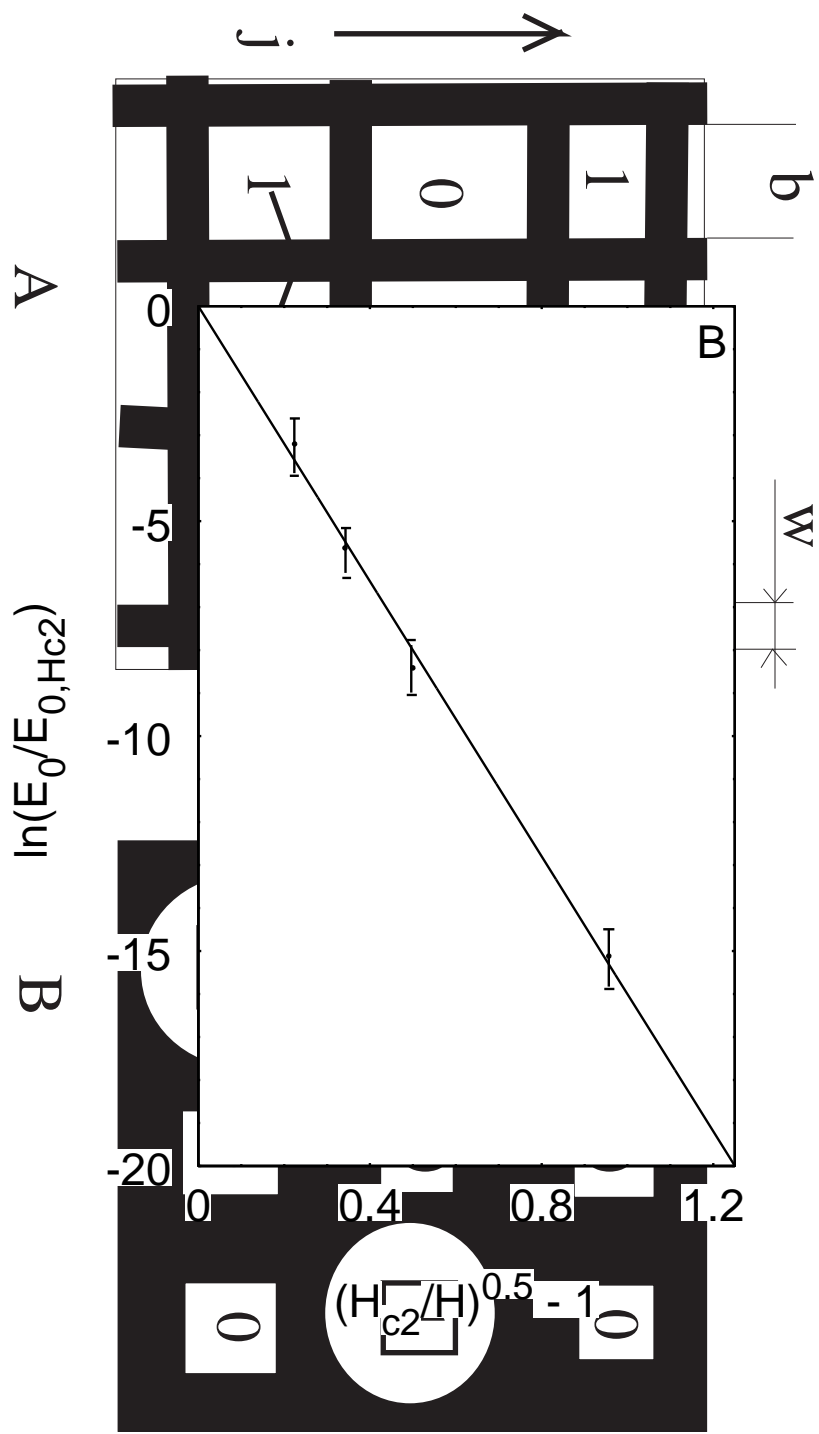
\end{document}